# Dynamic Evaluation of Power Distribution Lines for Determining the Number of Repair Teams and Prioritizing the Resolution of Faults Caused by High-Impact Low-Probability Events

Hamidreza Sharifi Moghaddam[1], Reza Dashti[2], Abolfazl Ahmadi[3]

1. Master of Science Candidate, Iran University of Science and Technology 2. Assistant Professor, Iran University of Science and Technology 3. Assistant Professor, Iran University of Science and Technology

**Abstract:** In order to increase the resilience of distribution systems against high-impact low-probability (HILP) events, it is important to prioritize the damaged assets so that the lost loads, especially critical and important loads, can be restored faster. In addition, correctly predicting the number of repair teams during critical times contributes to restoring the network to the initial resilience level. For this reason, this paper discusses the prioritization of electricity supply lines for evaluating the number of required repair teams. To this end, the economic value of distribution system lines has been considered as a criterion representing the sensitivity of the network to hurricanes. The modeling is based on value, in which the load value, failure probability of the poles, fragility curve, duration of line repair by the maintenance team, and the topology factor have been considered. This is so that the significance of the demand side, the failure extent and accessibility of the lines, the importance of time, and the network configuration are considered. The results provide a list of line priority for fault resolution, in which the topology factor has a larger effect. The number of repair teams required to restore critical and important loads is determined from this model. This modeling has been tested on an IEEE 33-bus network.

1. Introduction

The continuous climate change and global warming during recent years have exhibited their effect on climate HILP events. The events that were previously considered as low-probability are currently a potential threat for the provision of services and facilities to the public. In recent years, tornadoes, hurricanes, floods, and other climate HILP events have caused economic damage to societies, and with the current state, they are expected to occur more frequently and with larger financial damage and more casualties [1].

Nowadays, electrical energy is known as a clean and always-available energy, the ease of transmission and distribution and the high reliability of which have made it an inseparable part of mankind's life. Moreover, the dependence of industries and critical service centers that are directly related to public comfort and security on this energy has turned it into an irreplaceable pillar of human civilization. However, in recent decades, continuous access to this energy has been compromised by climate HILP events and human threats. As a result, studies concerning the

---

Author E-mail: rdashti@iust.ac.ir

assurance of continuous electrical energy supply to customers have been placed on researchers' agenda.

It has been decades that power system reliability follows a probabilistic approach toward system performance during a specific time period. These approaches relate to low-impact/high-probability disasters, such as equipment failure, repairs, storage and spare parts. Recent climate events show that one cannot prevent the grave consequences of HILP events using the existing tools and methods. Therefore, it is necessary to define a separate topic to examine the conditions of power systems in case of HILP events [1]. Investigating the solutions for reducing the vulnerability and reversibility of power systems against such events fall in the category of power system resilience topics. Ref. [2] presents a state-of-the-art review of existing research on the study of grid resilience and examines it in 4 sections: 1. Studying the words and phrases used in resilience 2. Presenting a resilience framework in network studies 3. Examining a number of resilience evaluation methods and quantitative indexes 4. Implementing resilience strategies

Previous studies, including [3], divide the actions in the field of power systems resilience into three main parts: 1. Reinforcement measures 2. Corrective measures and emergency response 3. Damage evaluation and restoration the studies performed in this field are discussed in the following.

1- **Reinforcement measures:** The US Department of Energy defines reinforcement as physical changes in the infrastructure that make the network less vulnerable to damage due to hurricane, floods, or airborne chips [4]. Among the preventive measures before the event that reinforce the system against events are vegetation, moving substations underground or up high, and using high-quality materials in the provision of accessories and tools. Many studies in power network reinforcement, such as [5], have focused on transmission network reinforcement considering natural hazards or terrorist attacks. Ref. [6] has proposed an improvement in flexibility by adding RCSs and reinforcing a number of the lines. Ref. [7] has proposed a three-level optimization technique involving the following: 1. Minimizing the investment cost to reinforce the lines 2. Selecting the worst-case scenario and finding the damaged distribution lines 3. Minimizing the load shedding cost. A dynamic network reconfiguration methodology with considering time-varying weather conditions is proposed, which the main goal is to minimize outage risk [8].

2- **Corrective measures and emergency response:** Immediately after the event, corrective measures and emergency response such as emergency load shedding, special protective systems, islanding, and switching are planned [1]. The majority of resilience studies relate to this part. A two-step robust optimization model has been proposed in [9] to change the distribution network structure according to the load uncertainty. Ref. [10] has proposed a restoration method with the division of the distribution network to microgrids with distributed generators (DGs). Ref. [11] presents an extended feeder restoration method for restoring critical loads using DERs, in which the maximum restoration of critical loads and optimal time are considered. Ref. [12] has adopted a self-healing resilience strategy and has examined the islanding of the parts in a distribution network based on the Benders decomposition method. Ref. [13] points out the significance of DGs in improving network resilience and uses this potential to form microgrids and reconfigure the network topology. Ref. [14] presents an improved distribution system restoration algorithm with microgrids. This algorithm enhance the survivability of out of service loads due to extreme events, like natural disasters. An optimal strategy to restore maximum loads with minimal switching operations under single and multiple fault conditions is proposed. Ref. [15] proposes a new permutation-based model for power system restoration within an optimized flexible duration that available generator capability and load prioritization are considered.

3- **Damage evaluation and restoration:** The ultimate restoration involves evaluating the damage and performing repairs, commissioning microgrids, and organizing repair teams and mobile transformers. Experience has shown that poet-event corrective actions can restore only a part of the lost loads to the network and that, after these actions, crisis management and planning related to commissioning microgrids and to repair teams are important. Moreover, after the occurrence of an event, it is possible for the damage to intensify and for unidentified damages to appear during the restoration process. Ref. [16] presents a comprehensive analysis of the role of microgrids in resilience improvement through the formation of self-sustainable microgrids and networked microgrids, in which the resilience of communication networks and the microgrid components. Ref. [17] combines economic-technical goals with resilience and reliability and installs storage and distributed generation to complement renewable generation and enable the microgrid to supply priority demands during stochastic islanding events. Ref. [18] addresses the effect of emergency diesel generators during crisis and shows that single emergency diesel generators configurations are only 80% likely to provide power for the duration of a two-week grid outage.

Ref. [19] presents an online spatial risk analysis that is expressed using a Severity Risk Index (SRI), is monitored and supported on-line, and can indicate risks in the process of forming in different parts of a power system under HILP accidents. Ref. [20] presents a time-dependent resilience criterion that examines several restoration prioritization plans. These resilience criteria have been used to evaluate the event severity reduction procedures, adaptation, or reinforcement strategies in power distribution systems against HILP disaster. Ref. [21] presents an Agent Based Modeling (ABM) approach to optimize the post-hurricane restoration procedure. It concludes that parameters such as the number of outages, repair time range, and number of staff can considerably affect the Estimated Time to Restoration (ETR) and that other parameters such as the location of work and movement speed have slight effects on ETR.

Ref. [22] proposes a system resilience reinforcement strategy after a HILP event according to the repair of damaged components and system operational dispatching. The repair tasks, unit output, and the switching plan are designed according to the results of damage evaluation. This reference proposes a mixed-integer optimization problem by combining the Vehicle Routing Problem (VRP) and DC load distribution model to reduce outage losses and rapidly store load supply resources.

Electrical engineers seek to reduce risks in reliability, and the measures taken are evaluated from an economic perspective. However, the planning model used in reliability cannot be used in resilience since the probability of HILP event is low but its consequences are grave. Therefore, we cannot seek for invulnerability in resilience since it is virtually impossible and also very costly. On the other hand, there are financial limitations in distribution companies, and the need for convincing managers to invest in this field makes things more difficult. However, distribution companies readily accept crisis resolution as an important responsibility.

The authors of [23] have presented a conceptual resilience curve and have stated the actions related to each part in the two classes of short-term and long-term actions in [24]. In this paper, this curve is developed in the form of Fig. 1 by highlighting the role of short-term resilience actions and the actions taken after fast restoration.

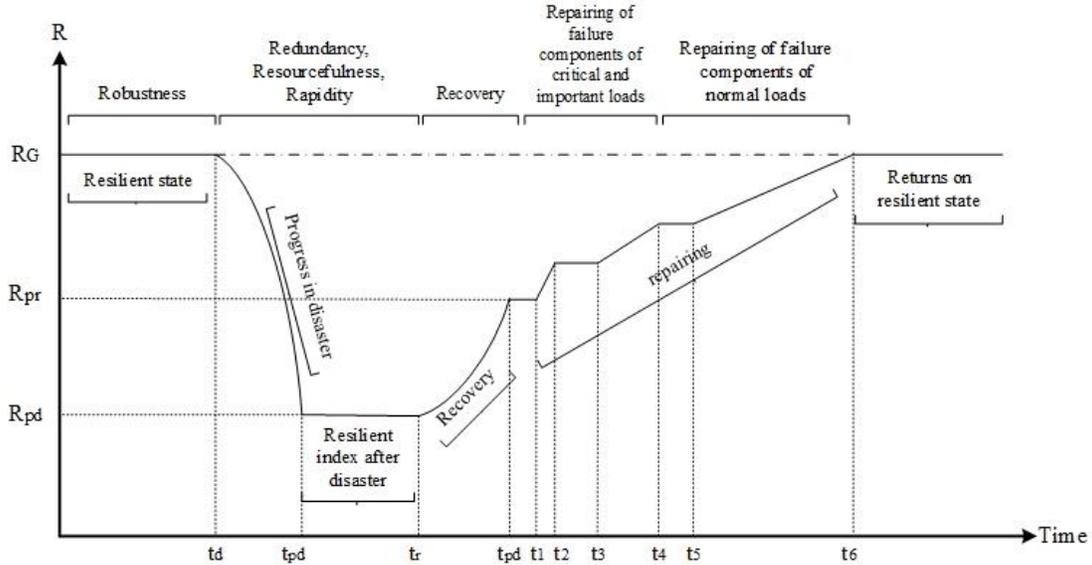

**Figure 1.** Resilience curve with a focus on final restoration (repairs)

Resilience definitions show that this index is a function of time. The studies operation range is the time period between $t_{pd}$ and $t_6$. In this situation, reducing the time intervals in each of the following is important:

- $t_{pd} - t_1$: Related to the duration for preparation, task assignment, dispatch, and the establishment of repair teams at the critical points of the network.
- $t_1 - t_2$: The duration of the final restoration and the reconnection of the critical loads, such as hospitals and food and medication depots, to the network by the repair team.
- $t_2 - t_3$: The identification and repair planning for transferring the repair teams toward important loads and loads with a higher failure potential.
- $t_3 - t_4$: The duration for the final restoration of important loads.
- $t_4 - t_5$: The duration for task assignment of the repair teams called, backup force, and existing teams for restoring normal loads such as residential areas.
- $t_5 - t_6$: This time interval includes the final restoration of normal loads, fault resolution of the overall network, and ensuring of the full network restoration and return to the initial resilience level of the network.

A decrease in the time intervals depends on factors such as the type and extent of the disaster, pre-determined plans and actions, created models, and the number of teams and their speed and skill. Considering the restoration time is important after the event and during damage evaluation and the continuation of restoration for managing climate HILP events. On the other hand, one responsibility of distribution companies is to attempt to minimize the consequences of damage to the customers by taking into account economic considerations. During reinforcement before the event and corrective actions after the event, one can neither make the system invulnerable via large expenditures nor restore all loads to the pre-event via corrective actions. This indicates the significance of assessing damages, repairs, and restoration after the corrective actions. This paper aims to investigate approaches for improving the restoration process. Actions during the restoration include inspecting assets, monitoring the consumption status, and determining new consumption. During a crisis, emergency generation must correctly enter the network, unnecessary consumption must be eliminated, and the existing power must be distributed in order of consumption priority depending on what percentage of consumption the available power is. After power is correctly managed, one must identify and prioritize for repairs the more critical and important consumptions. This paper specifically targets the dynamic evaluation of the distribution system lines to prioritize the lines, to repair the assets, restore the loads, and help the network to reach its initial resilience level. The modeling is based on value, and the load value, pole failure probability considering lifetime and using the fragility curve, duration of repair by the repair team, and topology factor have been considered, respectively, to consider the importance of the demand side, examine the accessibility of the lines, consider time, and emphasize the network configuration.

## 2. Proposed model

The modeling here is based on value, and the value of the lines is considered as a criterion representing the sensitivity of the network to any disaster (hurricane in this paper). The smaller the value assigned to the network lines, this sensitivity will fall to a lower degree. In other words, the lines will be assigned a higher resilience in the resilience ranking and a lower priority in the repair prioritization. This evaluation contains the major factors, which will be pointed out in the following.

### 2.1. Load value

The accurate evaluation of damage and outages indicates the graph of the existing crisis and the social and equipment damage that the network will face, a better estimation of which is of great help to the faster repair and restoration of the network. It must be noted that not all the gathered data have the same value. The data relating to critical loads and the assets fed by such loads have priority over other loads. The relationship proposed to compute the load value is in the form of (1).

$$v_i = L(i) \times 8760 \times LF(i) \times value_i \qquad (1)$$

In the above equation, $L$ is the load consumed by the bus connected to the line $i$, $LF$ is the load factor index, and $value_i$ is the value of the lost loads, in which the load importance is considered. This importance has been displayed in (2).

$$value_i = CRT \times voll_b \; ; \quad CRT = \begin{cases} critical\ load: 100 \\ important\ load: 10 \\ ordinary\ load: 1 \end{cases} \qquad (2)$$

The large number of examinations indicates the requirement for resilience in important support systems for critical loads. Therefore, this importance is modeled using the factor $value$. In this modeling, loads such as hospitals and food and medication depot are considered as critical loads, traffic lights are considered as important loads, and other loads are considered as regular loads.

## 2.2. Failure probability of poles at a particular wind speed

The hypothetical event in this paper is considered to be hurricane. Therefore, the failure probability of the poles is modeled with respect to changes in the wind speed. The poles have different lifetimes. This difference in lifetime manifests in the reliability-based failure probability, the tolerance threshold at different wind speeds, and the fracture limit of the poles. Using these values, one can plot the fracture curve of each pole.

$$FC(p_0, v_{th}, v_{max}) \tag{3}$$

Where $p_0$ is the failure probability considering reliability and increases with an increase in lifetime. $v_{th}$ is the wind speed corresponding to the tolerance threshold of the pole and increases with an increase in lifetime. Moreover, $v_{max}$ represents the maximum wind speed that the pole can tolerate and decreases with an increase in lifetime.

Given the above explanations, the network poles, which have different lifetimes, are classified into k- classes or intervals in this research. The time intervals, which are the classification criterion, are determined using experimental studies of the vulnerability of the poles to previous hazards. Each interval has different $FC$s, as shown in Fig. 2.

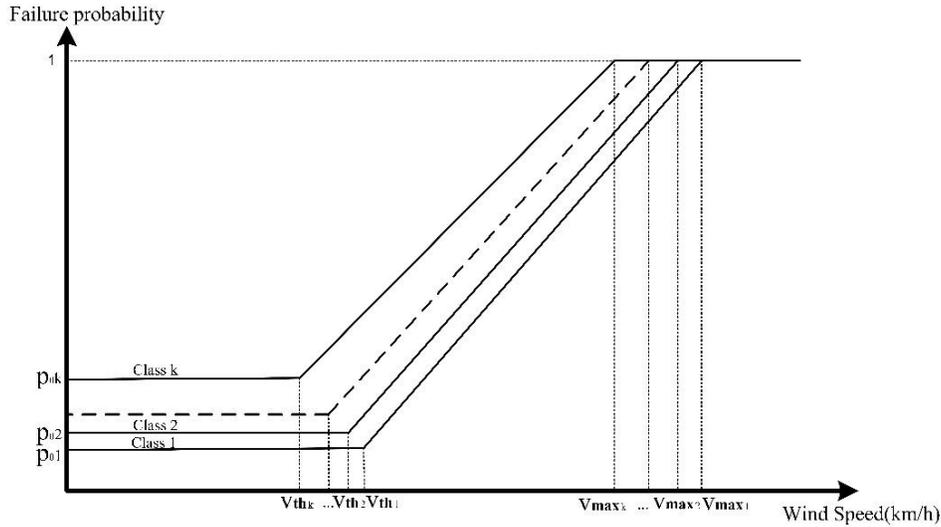

**Figure 2.** Failure probability of poles based on lifetime

In this figure, k- classes has a longer life and first class has a shorter lifetime.
If the classification of assets is in the form of Table 1:

**Table 1.** Classification of assets with respect to their lifetime

|  | $R_1$<br>0 - $LFT_1$ | $R_2$<br>$LFT_1$ - $LFT_2$ | $R_3$<br>$LFT_2$ - $LFT_3$ | ... | $R_K$<br>$LFT_{K-1}$ - $LFT_K$ |
|---|---|---|---|---|---|
| $p_0$ | $p_{0_1}$ | $p_{0_2}$ | $p_{0_3}$ | ... | $p_{0_K}$ |
| $v_{th}$ | $v_{th_1}$ | $v_{th_2}$ | $v_{th_3}$ | ... | $v_{th_K}$ |
| $v_{max}$ | $v_{max_1}$ | $v_{max_2}$ | $v_{max_3}$ | ... | $v_{max_K}$ |
| $n_i$ | $n_1$ | $n_2$ | $n_3$ | ... | $n_K$ |

Where $LFT_K$ is the pole life, and $n_i$ is the number of poles in each class. Then;

$$p_{0_1} < p_{0_2} < p_{0_3} < \cdots < p_{0_K} \tag{4}$$
$$v_{th_1} < v_{th_2} < v_{th_3} < \cdots < v_{th_K} \tag{5}$$

$$v_{max_1} > v_{max_2} > v_{max_3} > \cdots > v_{max_K} \tag{6}$$

## 2.3. Number of damaged poles

In addition to the discussions in section 2.2, the wind speed ($v_{real}$) is required to obtain the failure probability of the poles with respect to changes in the wind speed. For each $R_i$ class, the graph of $FC$ will be as in Fig. 3.

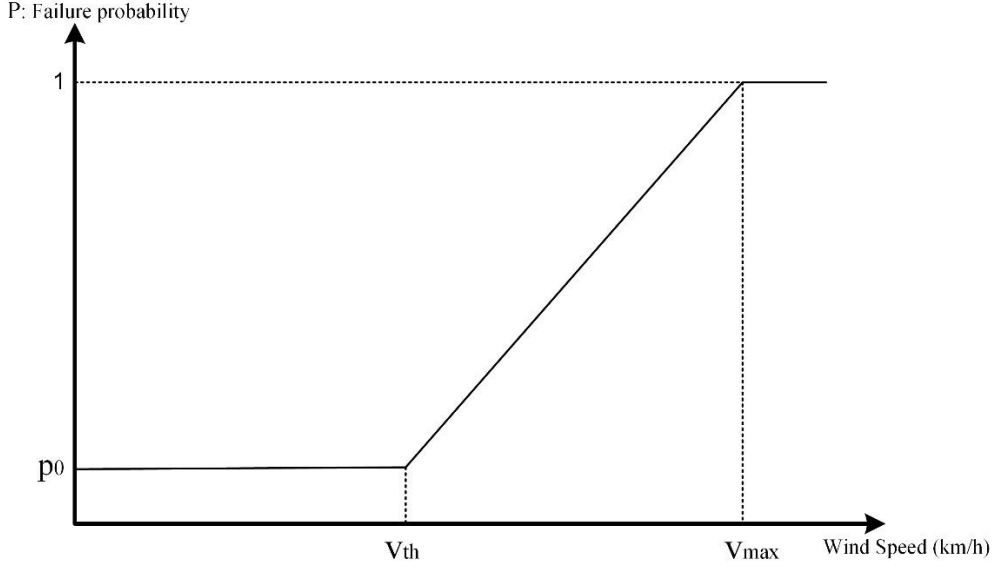

**Figure 3.** Reliability-based fragility curve of the poles

In this case, the failure probability of the poles is obtained from (8).

$$\forall i \, ; \, q_{R_i} = \begin{cases} p_{0_i} & ; \, v_{real} < v_{th_i} \\ m_i(v_{real} - v_{th_i}) + p_{0_i} & ; \, v_{th_i} < v_{real} < v_{max_i} \\ 1 & ; \, v_{max_i} < v_{real} \end{cases} \tag{8}$$

Where;

$$\forall i \, ; \, m_i = \frac{1 - p_{0_i}}{v_{max_i} - v_{th_i}} \tag{9}$$

As such, the number of damaged poles in each $R_i$ class ($b_{R_i}$) is determined using (10).

$$\forall i \, ; \, b_{R_i} = q_{R_i} . n_i \tag{10}$$

## 2.4. Number of damaged poles in a line

After the event and according to the relationships examined so far, one can obtain only the number of lines requiring repair in each class (lifetime). However, there are no data regarding the number of damaged poles in each line. Hence, (11) is used to determine the number of damaged poles in each line. Given that the pole lifetime is not important during a crisis, the pole lifetime has been used in this relationship, but it has not been considered in determining the ultimate number of damaged poles.

$$bt_j = \sum_{i=1}^{4} \frac{b_{R_i}}{n_i} . n_{c_{ji}} \tag{11}$$

Where $bt_j$ is the total number of damaged poles in line $j$. Also, $b_{R_i}$ is the number of damaged poles in the class $R_i$, $n_i$ is the total number of poles in the class i, and $n_{c_{ji}}$ is the total number of poles in each line in the class i.
The above equation is used, and the time for restoring each line to the network is considered in the model, as shown in (12).

$$t_{rep_j} = bt_j \times t_{rep_{av}} \qquad (12)$$

Where $t_{rep_{av}}$ is the average time-to-repair of each pole, considered to be 4 hours.

### 2.5. Dynamic value of the load
The value of the load during a crisis is known as the dynamic value of the load. It is computed according to (13).

$$v_{dyn_i} = (L(i) \times 8760 \times LF(i) \times value) \cdot t_{rep_i} \qquad (13)$$

Where $t_{rep_i}$ is duration required for the repair team to return each line to the network and is obtained from (14).

$$t_{rep_i} = \sum_{c=1}^{b} t_{rep_c} \qquad (14)$$

Where $t_{rep_c}$ is the repair duration of each pole in the line, and $b$ is the number of damaged poles in the line that need repair.

### 2.6. Topology
It can be claimed that the most fundamental indicator in evaluating the lines in this paper is their position within the network. According to the modeling, the upstream lines possess also the value of all their respective downstream lines. Hence, the topology of the network is considerably effective in this sense.

### 2.7. Dynamic value of a line
The values of the lines is considered as a criterion representing the sensitivity of the network to any risk. The smaller the value assigned to the network lines, the lower the degree the sensitivity will fall. In other words, the lines will be assigned a higher resilience in the resilience ranking and a lower priority in the repair prioritization. The value of each line is obtained from the proposed formula (15).

$$v_{l_{dyn}}(i) = v_{dyn_i} + \sum_{j \in A} v_{dyn_j} \qquad (15)$$

Where A is the set of lines that become unavailable in case of a disconnection in the line i and whose life and value depends on the availability of the line i. In other words, the set of lines A represents the effect of the network topology, which has the most fundamental effect on the evaluation. Since the prioritization is done during crisis and, hence, must be performed in a short time, and since the lines are compared with each other in this research, the inherent value of the line, which is related to the investment cost for building the lines, is not considered in the numerical computations that follow. The value computation model for each line is displayed in Fig. 4.

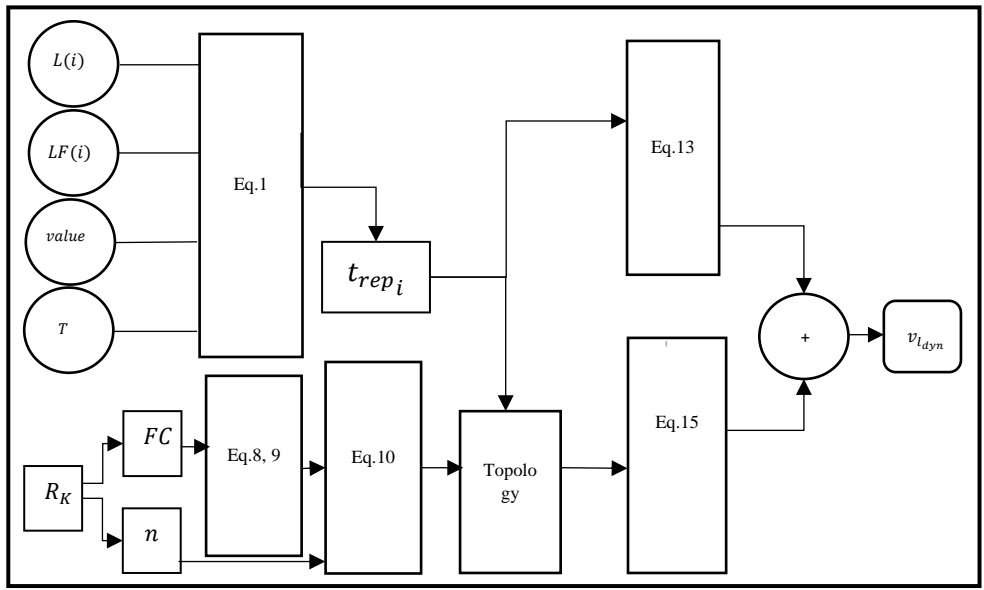

**Figure 4.** Load value computation model

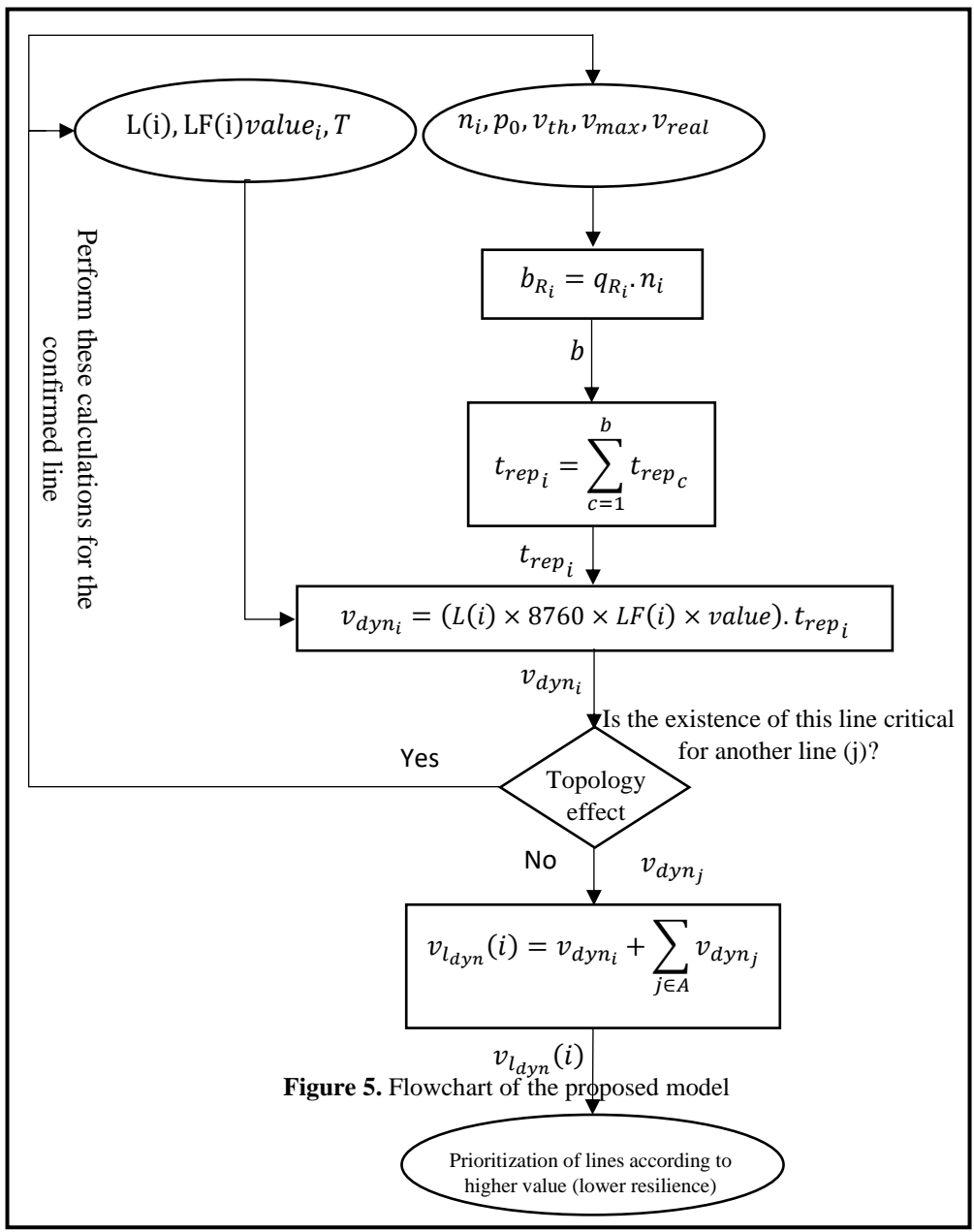

**Figure 5.** Flowchart of the proposed model

## 3. Numerical Studies
The 33-bus network of Fig. 6 was considered for evaluating and confirming the proposed model.

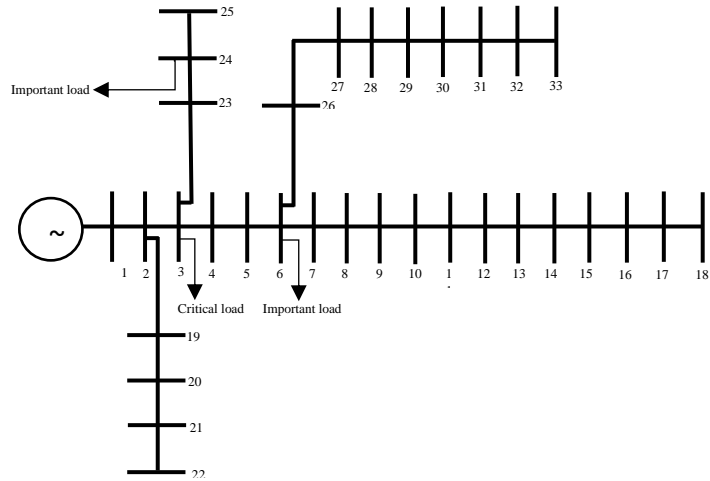

**Figure 6.** IEEE 30-bus standard network

The specifications of the actual power received by the first feeder were presented in Table 2:

**Table 2.** Specifications of the power received by the buses

| Sending bus | Receiving bus | $L$ | Sending bus | Receiving bus | $L$ | Sending bus | Receiving bus | $L$ |
|---|---|---|---|---|---|---|---|---|
| 1 | 2 | 100 | 12 | 13 | 60 | 23 | 24 | 420 |
| 2 | 3 | 90 | 13 | 14 | 120 | 24 | 25 | 420 |
| 3 | 4 | 120 | 14 | 15 | 60 | 6 | 26 | 60 |
| 4 | 5 | 60 | 15 | 16 | 60 | 26 | 27 | 60 |
| 5 | 6 | 60 | 16 | 17 | 60 | 27 | 28 | 60 |
| 6 | 7 | 200 | 17 | 18 | 90 | 28 | 29 | 120 |
| 7 | 8 | 200 | 2 | 19 | 90 | 29 | 30 | 200 |
| 8 | 9 | 60 | 19 | 20 | 90 | 30 | 31 | 150 |
| 9 | 10 | 60 | 20 | 21 | 90 | 31 | 32 | 210 |
| 10 | 11 | 45 | 21 | 22 | 90 | 32 | 33 | 60 |
| 11 | 12 | 60 | 3 | 23 | 90 | - | - | - |

Other specifications required by the 33 evaluated lines are displayed in Table 3.

**Table 3.** Specifications of the network lines

| Num.l | $l_f$ | $value_i$ | Num.l | $l_f$ | $value_i$ | Num.l | $l_f$ | $value_i$ |
|---|---|---|---|---|---|---|---|---|
| 1 | 0.8 | 3200 | 12 | 0.91 | 3200 | 23 | 0.91 | 3200 |
| 2 | 0.9 | 3200 | 13 | 0.8 | 3200 | 24 | 0.8 | 32000 |
| 3 | 0.85 | 3200 | 14 | 0.9 | 3200 | 25 | 0.9 | 3200 |
| 4 | 0.88 | 320000 | 15 | 0.85 | 3200 | 26 | 0.85 | 3200 |
| 5 | 0.89 | 3200 | 16 | 0.88 | 3200 | 27 | 0.88 | 3200 |
| 6 | 0.91 | 32000 | 17 | 0.89 | 3200 | 28 | 0.89 | 3200 |
| 7 | 0.8 | 3200 | 18 | 0.8 | 3200 | 29 | 0.91 | 3200 |
| 8 | 0.9 | 3200 | 19 | 0.9 | 3200 | 30 | 0.8 | 3200 |
| 9 | 0.85 | 3200 | 20 | 0.85 | 3200 | 31 | 0.9 | 3200 |
| 10 | 0.88 | 3200 | 21 | 0.88 | 3200 | 32 | 0.85 | 3200 |
| 11 | 0.89 | 3200 | 22 | 0.89 | 3200 | 33 | 0.88 | 3200 |

According to the data obtained from power distribution companies, 3 to 10 poles are installed between every two buses. In this network. The total number of poles was considered to be 240, and the normal distribution function was used to distribute these poles in the four classes mentioned. The number of poles is shown in Figs. 7, in which each column represents the number of poles and the horizontal axis displays the pole lifetime. The pole data are in the form of Table 4. In this table, the poles are divided into 4 clsses (in terms of lifetime) the number of poles in each of which is obtained from the normal distribution. The $p_0$s, which are the failure probability considering reliability, are specified in the table for each class. Furthermore, $v_{th}$ and $v_{max}$ have been specified for each class in the table. All these data are model inputs and have been obtained from experimental data.

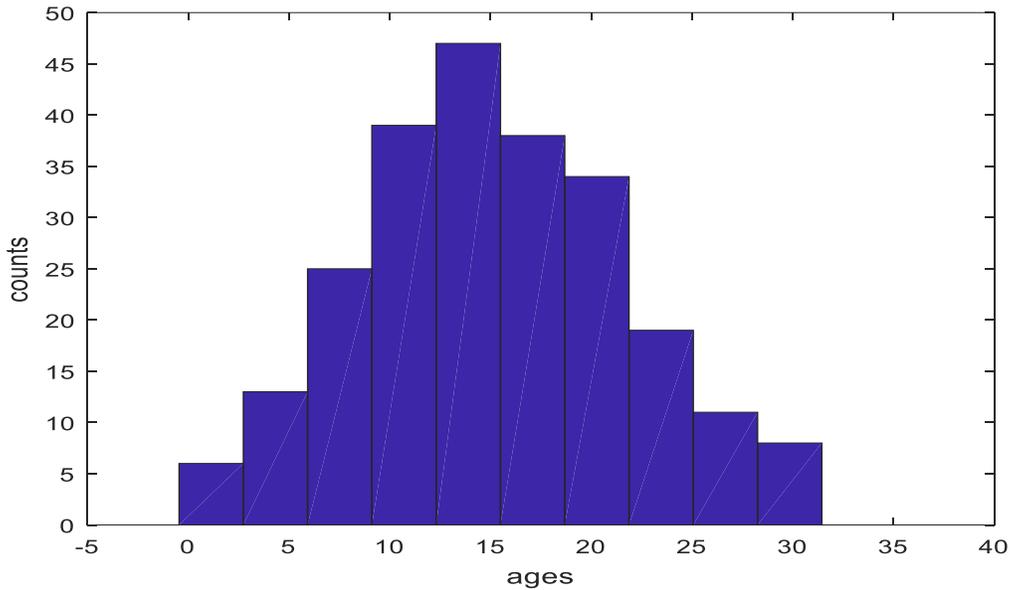

**Figure 7.** Pole distribution in the 33-bus system

**Table 4.** Pole data in the 33-bus system

| Poles | Pole lifetime (year) | Number of poles | $p_0$ | $v_{th}(\frac{km}{h})$ | $v_{max}(km/h)$ |
|---|---|---|---|---|---|
| First class | 0-5 | 15 | 0.05 | 110 | 150 |
| Second class | 5-15 | 106 | 0.07 | 100 | 140 |
| Third class | 25-15 | 98 | 0.09 | 95 | 120 |
| Fourth class | Above 25 | 21 | 0.11 | 90 | 115 |

The poles in each line are divided into four classes. These four classes have different parameters that's shown in Fig. 8.

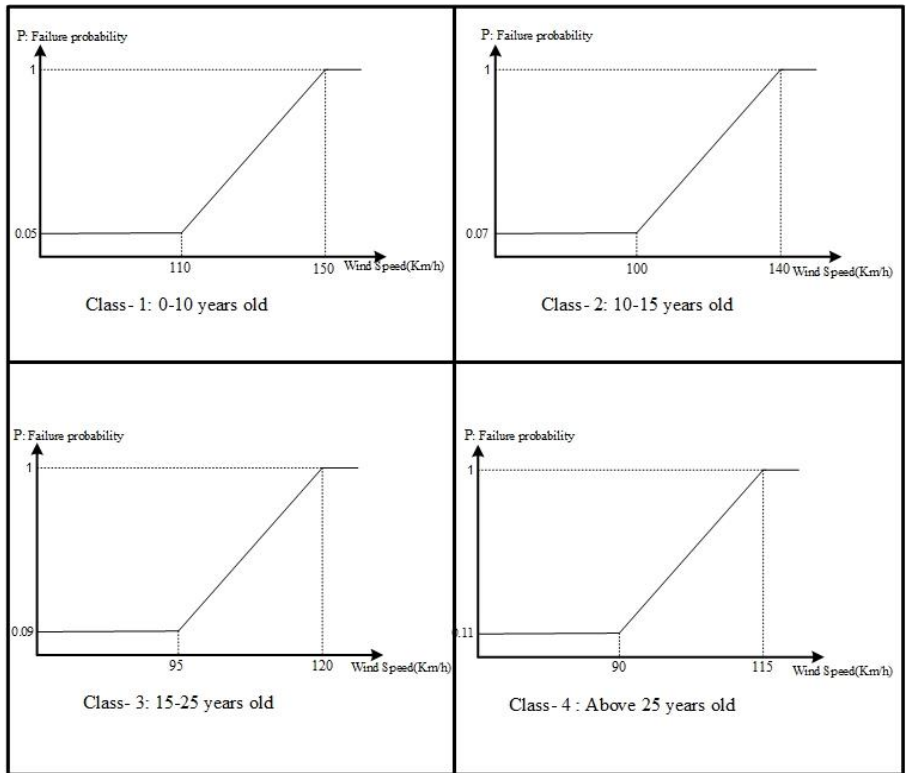

**Figure 8.** Fragility curve of poles based on lifetime

The number of poles in each line are shown divided into classes in Table 5. This information has been considered according to experimental data. For example, there are four poles in the second line, with one pole in the class 2 and three poles in the class 4.

**Table 5.** Number of poles in each line based on lifetime

| line | Total number of Pole | 0-5 | 5-15 | 15-25 | Over 25 | line | Total number of Pole | 0-5 | 5-15 | 15-25 | Over 25 |
|---|---|---|---|---|---|---|---|---|---|---|---|
| 1 | 4 | - | 1 | - | 3 | 18 | 9 | - | 2 | 7 | - |
| 2 | 4 | - | 1 | - | 3 | 19 | 10 | - | 8 | 2 | - |
| 3 | 6 | - | 1 | 1 | 4 | 20 | 8 | 1 | 1 | 6 | - |
| 4 | 7 | - | 1 | 1 | 5 | 21 | 8 | 1 | 1 | 6 | - |
| 5 | 6 | - | 1 | 1 | 4 | 22 | 7 | 1 | 6 | - | - |
| 6 | 6 | 1 | - | 4 | 1 | 23 | 10 | - | 1 | 9 | - |
| 7 | 7 | - | 2 | 5 | - | 24 | 7 | 1 | 1 | 5 | - |
| 8 | 7 | 1 | 6 | - | 1 | 25 | 8 | - | 6 | 2 | - |
| 9 | 7 | - | 1 | 6 | - | 26 | 9 | 1 | 7 | 1 | - |
| 10 | 8 | - | 2 | 5 | - | 27 | 8 | 1 | 6 | 1 | - |
| 11 | 7 | - | 1 | 6 | - | 28 | 6 | - | 1 | 5 | - |
| 12 | 6 | - | 5 | 1 | - | 29 | 7 | 5 | 1 | 1 | - |
| 13 | 8 | 1 | 2 | 5 | - | 30 | 7 | - | 6 | 1 | - |
| 14 | 8 | - | 2 | 6 | - | 31 | 8 | - | 1 | 7 | - |
| 15 | 7 | - | 6 | 1 | - | 32 | 7 | 1 | 6 | - | - |
| 16 | 8 | - | 7 | 1 | - | 33 | 7 | - | 7 | - | - |
| 17 | 8 | - | 6 | 2 | - | - | - | - | - | - | - |

The failure probability of the poles with respect to wind speed was obtained as shown in Table 6. The number of damaged poles are shown according to (8), (9), and (10). In this relationship, decimals are rounded to the upper integer.

**Table 6.** Number of damaged poles in each class at various hurricane speeds

| $v_{real}(\frac{km}{h})$ | $q_{R_1}$ | $q_{R_2}$ | $q_{R_3}$ | $q_{R_4}$ |
|---|---|---|---|---|
| 80 | 1 | 7 | 9 | 2 |
| 90 | 1 | 7 | 9 | 2 |
| 100 | 1 | 7 | 27 | 10 |
| 110 | 1 | 32 | 62 | 17 |
| 120 | 4 | 57 | 98 | 21 |
| 130 | 8 | 81 | 98 | 21 |
| 140 | 11 | 106 | 98 | 21 |
| 150 | 15 | 106 | 98 | 21 |

The above table is displayed in Fig. 9 in pictorial form

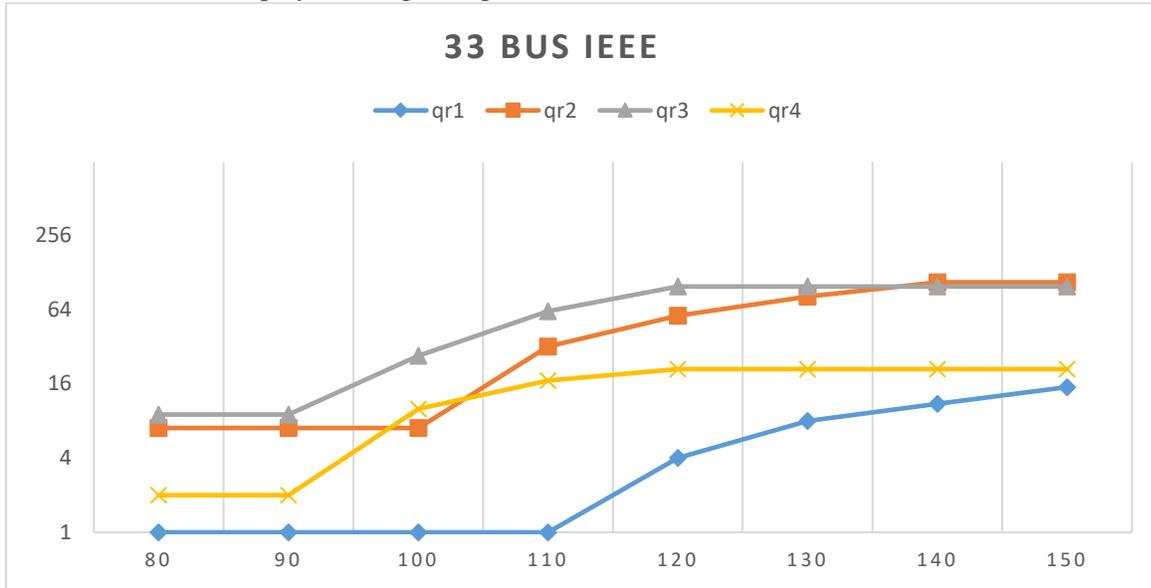

**Figure 9.** Number of damaged poles based on lifetime

As seen in the fig. 9, with an increase in the wind speed, the number of damaged poles in each class and, hence, in the network increases. For the rest of the calculations, the wind speed is assumed to be $v_{real} = 105 \ km/h$.

### 3.1. Calculation of the time-to-restoration of each line
Using the mentioned relationships and considering the repair time of pole to be 4 hours led to Table 7.

**Table 7.** Repair duration for each line

| Line.num | 1 | 2 | 3 | 4 | 5 | 6 | 7 | 8 | 9 | 10 | 11 |
|---|---|---|---|---|---|---|---|---|---|---|---|
| $bt_j$ | 2.1183 | 2.1183 | 3.2163 | 3.8603 | 3.2163 | 2.51 | 2.6426 | 1.7618 | 2.9103 | 2.6926 | 2.9103 |
| $t_{rep_j}$ | 8.4732 | 8.4732 | 12.8652 | 15.4412 | 12.8652 | 10.0400 | 10.5704 | 7.0472 | 11.6412 | 10.7704 | 11.6412 |
| Line.num | 12 | 13 | 14 | 15 | 16 | 17 | 18 | 19 | 20 | 21 | 22 |
| $bt_j$ | 1.3855 | 2.6926 | 3.0966 | 1.5718 | 1.7581 | 2.0258 | 3.5506 | 2.3984 | 2.9603 | 2.9603 | 1.1678 |
| $t_{rep_j}$ | 5.5420 | 10.7704 | 12.3864 | 6.2872 | 7.0324 | 8.1032 | 14.2024 | 9.5936 | 11.8412 | 11.8412 | 4.6712 |
| Line.num | 23 | 24 | 25 | 26 | 27 | 28 | 29 | 30 | 31 | 32 | 33 |
| $bt_j$ | 4.2723 | 2.5063 | 2.0258 | 1.8081 | 1.6218 | 2.4563 | 0.8903 | 1.5718 | 3.3643 | 1.1678 | 1.3041 |
| $t_{rep_j}$ | 17.0892 | 10.0252 | 8.1032 | 7.2324 | 6.4872 | 9.8252 | 3.5612 | 6.2872 | 13.4572 | 4.6712 | 5.2164 |

In the studied network, the exact locations of the damaged poles are determined using GIS. Moreover, the repair team is considered to be accommodated at the slack bus. Also, the time needed for the repair team to arrive is considered to be zero.

According to the mentioned relationships and considering the network topology, the simulation was performed using MATLAB software, and the results are shown in order of value in Table 8 and Fig. 10.

**Table 8.** Simulation results

| Evaluation rank | Value ($\times 10^{13}$) | Line Number | Evaluation rank | Value ($\times 10^{13}$) | Line Number | Evaluation rank | Value ($\times 10^{13}$) | Line Number |
|---|---|---|---|---|---|---|---|---|
| 1 | 3.6517 | 3 | 12 | 0.0727 | 11 | 23 | 0.0146 | 19 |
| 2 | 2.4646 | 1 | 13 | 0.0622 | 28 | 24 | 0.0134 | 30 |
| 3 | 2.4586 | 2 | 14 | 0.0571 | 14 | 25 | 0.0129 | 12 |
| 4 | 0.9217 | 23 | 15 | 0.0568 | 13 | 26 | 0.0115 | 25 |
| 5 | 0.6044 | 4 | 16 | 0.0502 | 31 | 27 | 0.0075 | 15 |
| 6 | 0.5236 | 24 | 17 | 0.0337 | 20 | 28 | 0.0072 | 16 |
| 7 | 0.4671 | 5 | 18 | 0.0270 | 8 | 29 | 0.0067 | 17 |
| 8 | 0.2582 | 6 | 19 | 0.0215 | 26 | 30 | 0.0049 | 29 |
| 9 | 0.1215 | 7 | 20 | 0.0210 | 18 | 31 | 0.0028 | 32 |
| 10 | 0.0899 | 9 | 21 | 0.0192 | 21 | 32 | 0.0009 | 22 |
| 11 | 0.0751 | 10 | 22 | 0.0180 | 27 | 33 | 0.0008 | 33 |

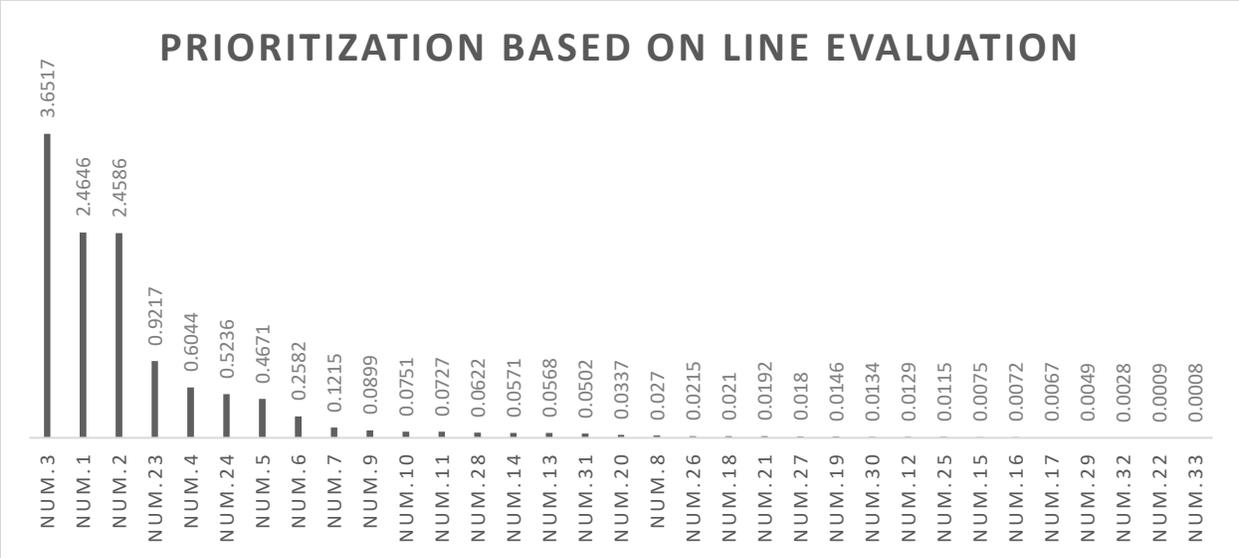

**Figure 10.** Dynamic evaluation results of the lines

To better understand the evaluation in prioritizing line repair, the evaluation heat map is shown in Fig. 11, in which priorities 1 to 11, 12 to 22, and 23 to 33 are displayed in red, orange, and green, respectively.

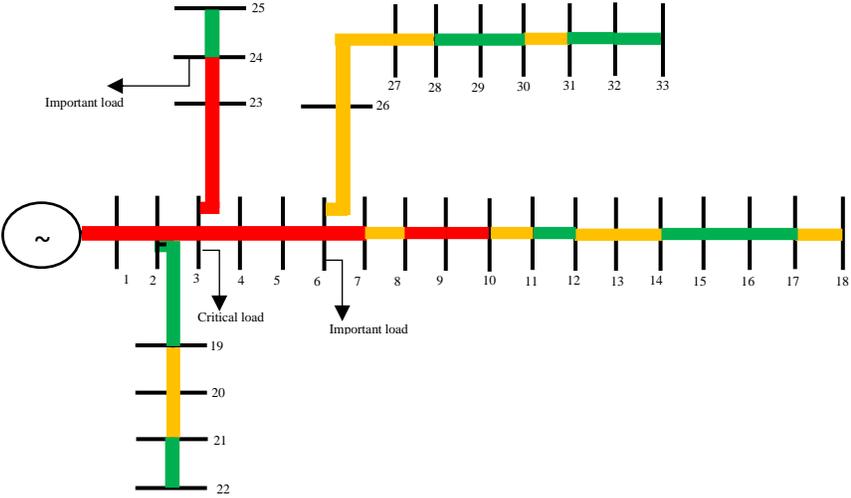

**Figure 11.** Heat map of line evaluation

Given that the number of the receiving bus represents the considered line number, the following issues are challenging: The network topology effect is clear in this heat map, such that the lines at the end of the network are at lower priorities, and the closer we get to the slack bus, the higher the value of the lines becomes. Lines 1 to 7 have the highest values due to their topological positions despite not supplying much load. Among these 7 lines, line 3 has the highest value due to the its sensitivity and ranks first in the evaluation ranking despite having a lower topological value compared to line 1 or 2. Another interesting issue is line 8, which ranks 18 in the evaluation ranking due to the lower vulnerability of its poles to hurricanes despite its importance due to its topology and location.

### 3.2. Calling the repair team

After the event, critical and important loads are the priority of distribution companies. When actions such as switching, diesel generators, etc. cannot restore these loads to the network, correctly organizing the repair team becomes important. In the following, we prioritize the repair of these lines according to the critical and important loads in the network under study.

To restore the loads 3, 6, and 24, the lines 3, 1, 2, 23, 24, 5, and 6 must be restored according to priority. Table 9 shows the repair duration and number of teams for repairing these lines. A total of 29 teams are required. These calculations assume that all the poles are exposed to the hurricane and the damage is maximal.

**Table 9.** Number of teams required to repair critical and important load

| Line.num | 3 | 1 | 2 | 23 | 4 | 24 | 5 | 6 |
|---|---|---|---|---|---|---|---|---|
| $bt_j$ | 3.2163 | 2.1183 | 2.1183 | 4.2723 | 3.8603 | 2.5063 | 3.2163 | 2.51 |
| Num.Repair Team | 4 | 3 | 3 | 5 | 4 | 3 | 4 | 3 |

### 4. Conclusion

Asset evaluation and, in this paper, line and pole evaluation are necessary for a fast response to HILP climatic events in restoration from the critical condition. Generally, the lines located at the end of the network ranked lowest in the prioritization, indicating the importance of the network topology. Most of the time, relatively disperse evaluation shows the necessity of attention to repair teams, which must move fast so as to speed up the load restoration process to reach the initial resilience level. The challenge for repair planning during and after unpredictable events can be observed clearly in the obtained heat map. In this paper, repair priority was given to critical and important loads, and 29 repair teams were required.